\documentclass[12pt,a4paper,final]{iopart}
\usepackage{amsthm,amsfonts,amssymb,amscd}
\usepackage{iopams}

\expandafter\let\csname equation*\endcsname\relax

\expandafter\let\csname endequation*\endcsname\relax

\usepackage{amsmath}
\usepackage{graphicx}
\usepackage{subfig}
\usepackage{tikz}
\usepackage{tikzit}

\tikzstyle{red dot}=[fill=red, draw=black, shape=circle]
\tikzstyle{green dot}=[fill={rgb,255: red,0; green,182; blue,0}, draw=black, shape=circle]
\tikzstyle{new style 0}=[fill=white, draw=black, shape=circle]
\tikzstyle{big box}=[fill=white, draw=black, shape=rectangle, minimum height=12cm]
\tikzstyle{new style 1}=[fill=white, draw=black, shape=circle, minimum height=12cm]
\tikzstyle{new style 2}=[fill=white, draw=black, shape=circle]

\tikzstyle{new edge style 0}=[draw=black, ->]
\tikzstyle{new edge style 1}=[-, draw=red]
\tikzstyle{new edge style 2}=[-, draw={rgb,255: red,0; green,0; blue,192}]

\usepackage[breaklinks=true,colorlinks=true,linkcolor=blue,urlcolor=blue,citecolor=blue]{hyperref}
\usepackage[utf8]{inputenc}	

\newcommand{\beq}{\begin{equation}}
\newcommand{\eeq}{\end{equation}}
\newcommand{\bea}{\begin{eqnarray}}
\newcommand{\eea}{\end{eqnarray}}

\begin{document}

\title[Free Variable Loop Equations for the 3-State Potts Model Coupled to 2D Gravity]{Free Variable Loop Equations for the 3-State Potts Model Coupled to 2D Gravity}

\author{Aravinth Kulanthaivelu}
\address{Rudolf Peierls Centre for Theoretical Physics, Clarendon Laboratory, Parks Road, University of Oxford, Oxford OX1 3PU, UK}
\ead{aravinth.kulanthaivelu@physics.ox.ac.uk}

\begin{abstract}
We revisit the long standing problem of the geometric free variable approach to computing the generating function for disk amplitudes in the matrix model formulation of the 3-state Potts model coupled to 2D discrete gravity. This method is a mild generalisation of Schwinger Dyson (loop-)equations approach, using graphical arguments to derive a single generating equation in non-commutative variables that contain the information of all loop equations. The method of constructing the generating equation is reviewed, we explain how a larger class of constraints than previously known can be constructed and we derive a closed set of equations that determine the spectral curve for the planar graph generating function.

\end{abstract}

\vspace{2pc}
\maketitle

\section{Introduction}

The Potts model is a well known statistical lattice model which one usually encounters shortly after learning about the Ising model. The Ising model is a spin system consisting of a graph endowed with spins on each vertex that can be in one of 2 different states, up (+) or down (-), with nearest-neighbour interactions between unequal spins. The partition function is then calculated as a sum over all possible configurations with associated Boltzmann weights. The 3-state Potts model is a simple generalisation where we allow for 3 different spin states with nearest-neighbour interactions between any neighbouring unequal spins. In this paper we will be considering dynamical lattice approach to the 3-state Potts model coupled to 2D Euclidean quantum gravity (see \cite{Francesco1995} for a review).

To couple a spin system to gravity we take the underlying graph that this model is defined on and make it a random element of a statistical ensemble. Then, by summing over all possible graphs in an ensemble average, we simulate the summation over all possible geometries that a gravitational path integral should entail. This process plays the discretised role of summing over all 2D geometries in the bosonic string. The continuum theories are described by a Liouville CFT coupled to the CFT representing the long-range excitations of the critical spin system which, for the 3-state Potts model, is the $(6,5)$ D-series Virasoro minimal model. These theories of Liouville gravity coupled to a $(p,q)$ minimal model are particularly simple string theories with a small number of observables, called minimal string theories \cite{Seiberg2005}.

To simplify things we define our model to be a dynamical triangulation, such that the graphs are all triangulated surfaces and the Potts spins lie at the center of each triangle. This is neatly captured by the following matrix model
\begin{equation}
\label{eq1.1}
    Z = \int\prod_i [dX_i] e^{-NS(\{X_i\})},
\end{equation}
where $i\in\{0,1,2\}$, the $X_i$ are $N\times N$ Hermitian matrices, the measure $\prod_i[dX_i]$ is the Lebesgue measure denoting component-wise integration,
\begin{equation}
    \label{eq1.2}
    [dX_i] = \prod_{1\leqslant a,b\leqslant N} d \text{Re}(X_i)^a_{\: \: b} \prod_{1\leqslant a,b \leqslant N} d \text{Im}(X_i)^a_{\: \: b},
\end{equation}
and the action is given by
\begin{equation}
    \label{eq1.3}
    S= \frac{1}{2(1+c-2c^2)}\text{Tr}[(1+c)\sum_i X_i^2 -2c\sum_{\langle ij\rangle} X_i X_j] - \frac{g}{3}\text{Tr}\sum_i X_i^3,
\end{equation}
where $\langle \rangle$ denotes distinct neighbouring spins.

\begin{figure}
    \centering
    \scalebox{0.8}{\input{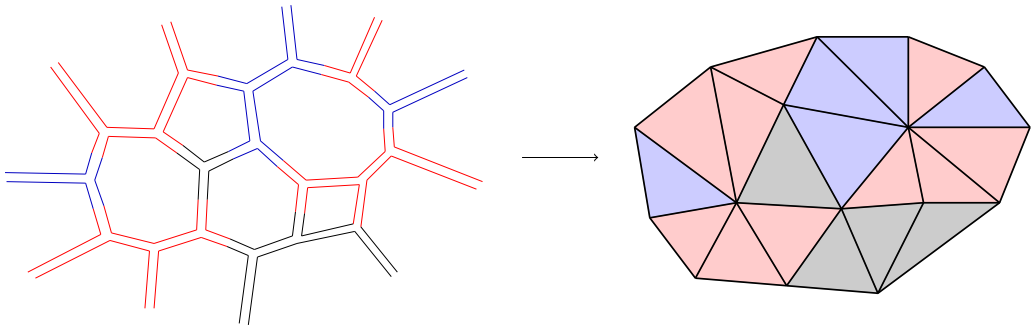}}
    \caption{Sample Feynman graph and dual triangulation, where different colours correspond to different spins.}
    \label{fig:PottsGraph}
\end{figure}

 The Feynman graph expansion of the above model consists of a series of ribbon graphs formed from trivalent vertices and propagators associated with the three different matrices. By forming the dual of these graphs, as demonstrated in Fig.\ref{fig:PottsGraph}, we generate a mesh giving the aforementioned triangulation. Furthermore, one can show that each graph carries an associated factor of $N^\chi$, where $\chi$ is the Euler characteristic of the graph. Therefore, in the large-N limit, our 2D graph expansion can be organised into an expansion in the topologies of graphs, with the leading order contribution coming from those diagrams that have spherical topology. We will exclusively work in this planar limit, but it is interesting to note that in many matrix models, once the planar observables are calculated, it is possible to reconstruct all subleading corrections in the topological expansion via a process called topological recursion \cite{Eynard2007}.
 
 The Potts matrix model was first written down for $q$ spins by Kazakov in \cite{Kazakov1988} and the solution was computed for the case of $q=1$ and $q=0$. For $q=2$ it reduces to the Ising model on a random lattice, which has been solved  \cite{Kazakov1986,Boulatov1987}. For the $q=3$ model, Daul and Zinn-Justin computed the one-loop function corresponding to the fully magnetised boundary condition on the spins and the associated critical exponent using the saddle-point method \cite{DAUL1995,Zinn-Justin1998}. For all allowed values of $q$ (see below)  Eynard and Bonnet \cite{Eynard1999} subsequently computed the one-loop function using the loop equation method. More recently, in \cite{Atkin2015,Atkin2016} a formalism was developed to compute a larger set of loop amplitudes corresponding to different boundary conditions, using a combination of the saddle-point method and loop equations. This formulation was illustrated by application to the $q=1,2,3$ Potts models and the known results reproduced, and new results for more general boundary conditions derived.
 
 In this paper we will derive the one-loop function of the 3-state Potts model using a geometric approach developed by Carroll, Ortiz and Wheeler in \cite{Carroll1996a}. The geometric method proposed therein involves the computation of a `generating equation' that describes the effect of removing a marked edge from a triangulation and the resulting diagrams. There is a correspondence between these moves and the generating equation, allowing it to be written down directly in terms of non-commutative variables after considering the effects of these moves. Once this has been computed, the next step is to derive a closed system of equations in terms of commuting variables. This technique is analogous to the method used in \cite{Douglas1995} in the context of the two-matrix model, but the generating equation provides a more general formulation. Computation of the one-loop function for the 3-state Potts model via this method was first addressed in \cite{Carroll1996a}, but they were unable to derive a closed set of equations. On account of \cite{Eynard1999,Atkin2016} we now know that the model is solvable, which we will demonstrate in this formalism.
 
 This article is organised as follows: in Section 2 we review the construction of the generating equation and derive it for the 3-state Potts model, in Section 3 we explain how loop equation constraints for the fixed spin one-loop function can be constructed from the generating equations and derive a closed system of equations for this disk amplitude. We then outline their solution in Section 4, where we write down the algebraic equation for the one-loop function. Finally, we discuss the implications of this result in Section 5 and provide an appendix illustrating the relationship between this method and direct reparameterisation via the matrix integral \eqref{eq1.1}.

\section{The Generating Equation}

\subsection{Pure Gravity}

To motivate the geometric approach used to study the 3-state Potts model coupled to gravity, we first consider the case of pure gravity. The discretised theory is given by the following matrix integral,
\begin{equation}
    \label{eq2.1}
    Z = \int [dX] \exp{\bigg(-N\text{Tr}\bigg[\frac{1}{2} X^2 - \frac{g}{3} X^3\bigg]\bigg)},
\end{equation}
where we now only integrate over a single $N\times N$ Hermitian matrix, $X$. 

We wish to compute the generating function for discretised surfaces with a single boundary, the disk amplitude. This is given by
\begin{equation}
    \label{eq2.2}
    \Phi (x,g) = \sum_{n=0}^{\infty}p_n(g)x^n = \frac{1}{N} \langle \text{Tr} \frac{1}{1-xX}\rangle
\end{equation}
where 
\begin{equation}
    \label{eq2.3}
    p_n(g) = \frac{1}{N} \langle \text{Tr} X^n \rangle .
\end{equation}
This can also be expressed in terms of $\mathcal{N}(k;n)$, the number of planar triangulations of a disk with a length $k$ boundary and $n$ triangles:
\begin{equation}
    \label{eq2.4}
    \Phi(x,g) = \sum_{k,n=0}^\infty \mathcal{N}(k;n)g^n x^k,
\end{equation}
To compute this object, we follow the prescription developed by Tutte \cite{Tutte1962} and derive recursion relations for the number of triangulations, which leads to a generating equation for $\Phi(x,g)$. Starting from a given triangulation with $n$ vertices and a $k$ length boundary we remove a marked edge, resulting in a number of smaller triangulations. The marked edge of the original triangulation can be connected in one of two ways:
\begin{enumerate}
    \item to a triangle on the boundary. In this case removing the marked edge results in a new triangulation which has one fewer triangles and one more boundary edge.
    \item to another boundary edge. In this case removing the marked edge removes the other edge as well and the triangulation is split into two triangulations with the same number of triangles but two fewer boundary edges.
\end{enumerate}
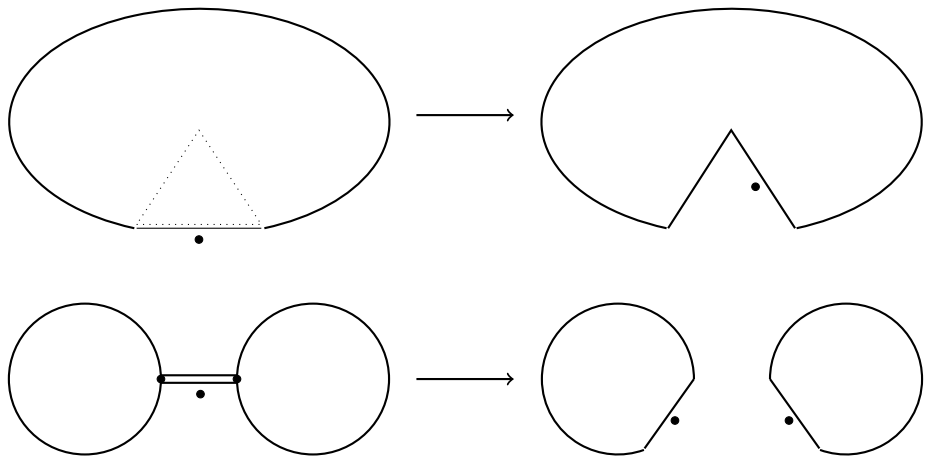
\begin{figure}
\centering
\begin{tikzpicture}
\draw[thick] (-3,0) arc (-70:250:2.5cm and 1.5cm);
\draw[dotted] (-4.68,0.05)--(-3.04,0.05) -- (-3.86,1.3) -- (-4.68,0.05);
\draw (-4.68,0) -- (-3.04,0);
\draw [thick,->] (-1,1.5) -- (0.28,1.5);
\draw[thick] (4,0) arc (-70:250:2.5cm and 1.5cm);
\draw[thick] (3.98,0) -- (3.14,1.3) -- (2.31,0);
\draw[fill] (-3.86,-0.15) circle[radius=0.05];
\draw[fill] (3.46,0.55) circle[radius=0.05];
\draw[thick] (-5.36,-2) circle (1cm);
\draw[thick] (-2.36,-2) circle (1cm);
\draw[fill] (-4.36,-2) circle[radius=0.05];
\draw[fill] (-3.36,-2) circle[radius=0.05];
\draw[thick] (-4.36,-2.05) -- (-3.36,-2.05);
\draw[thick] (-4.36,-1.95) -- (-3.36,-1.95);
\draw[fill] (-3.84,-2.2) circle[radius=0.05];
\draw [thick,->] (-1,-2) -- (0.28,-2);
\draw[thick] (3.65,-2) arc (180:-110:1cm and 1cm);
\draw[thick] (2.65,-2) arc (0:290:1cm and 1cm);
\draw[thick] (2,-2.92)--(2.65,-2);
\draw[thick] (4.3,-2.92) -- (3.65,-2);
\draw[fill] (2.4,-2.55) circle[radius=0.05];
\draw[fill] (3.9,-2.55) circle[radius=0.05];
\end{tikzpicture}
\caption{Result of removing an edge connected to a triangle (top), or another edge (bottom).}
\label{PureDecomp}
\end{figure}
To make these moves precise, a convention should be adopted regarding the resulting marked edge. We follow the convention adopted in \cite{Carroll1996a}; in case (i), once the original edge is removed we mark the edge oriented counterclockwise to the original removed edge, and in case (ii) we mark the two edges adjacent to those which were removed. This is illustrated in Fig.\ref{PureDecomp}. The recursion relation for the number of triangulations is then given by
\begin{equation}
    \label{eq2.5}
    \mathcal{N}(k;n) = \mathcal{N}(k+1;n-1) + \sum_{l=0}^{k-2}\sum_{m=0}^n \mathcal{N}(i;m)\mathcal{N}(k-2-l;n-m).
\end{equation}
The first term on the right hand side corresponds to case (i), where we have one fewer triangles and one more boundary edge, and the second term corresponds to case (ii), where we sum over two triangulations with $n$ triangles and $k-2$ boundary edges between them. To completely determine the number of triangulations this recursion relation needs to be supplemented with a boundary condition, which we take to be $\mathcal{N}(0;n)=\delta_{n,0}$, i.e. that there is only one trivial triangulation.

We can connect this to the generating function by simply performing the sum in \eqref{eq2.4}. The final expression for the generating function, which we call the generating equation, is then
\begin{equation}
    \label{eq2.6}
    \Phi(x,g) = 1 + \frac{g}{x^2}(\Phi(x,g)-1- p_1(g)x) +x^2\Phi^2(x,g).
\end{equation}
The first term on the right hand side corresponds to the boundary condition. The second term corresponds to case (i), where we subtract terms coming from triangulations with one and two boundary links because a minimum of two boundary edges are required if the original marked edge is connected to a triangle.  The last term corresponds to case (ii). This expression can be written more compactly by introducing a special derivative operator, $\Delta$, associated with the removal of a boundary triangle in case (i). This operator is defined to remove a single power of $x$ in the polynomial $x^k$, and annihilates any term independent of $x$: 
\begin{equation}
    \label{eq2.7}
    \Delta\sum_{k=0}^\infty a_k x^k = \sum_{k=1}^{\infty}a_k x^{k-1}.
\end{equation}
The generating equation is thus given by
\begin{equation}
    \label{eq2.8}
    \Phi(x,g) = 1 + gx\Delta^2\Phi(x,g) +x^2\Phi^2(x,g).
\end{equation}
Therefore we are able to identify the basic moves in the decomposition of a general triangulation with a precise operation. The second term in \eqref{eq2.8}, representing case (i), comes with a factor of $g$ and an operation $x\Delta^2$ acting on the generating function. The former corresponds to the weight of the removed triangle, whilst the latter corresponds to introducing two new edges, where the new marked edge is labelled by the fugacity $x$. The procedure given by case (ii) is associated with a factor of $x$ for each new marked edge and two factors of $\Phi$ for the new triangulations that remain after removing the boundary edges. 

Once the generating equation is constructed, a series of constraints for the generating function can be derived by applying a sequence of the derivative operators to it. We call these constraints loop equations, and they can be shown to be identical to the Schwinger-Dyson equations of the underlying matrix model \eqref{eq2.1}. Now that we have outlined the construction of the generating function and the translation between the basic geometric moves and associated operations on the generating function, we will extend this to the 3-state Potts model coupled to 2D gravity.

\subsection{3-State Potts Model}

The matrix model formulation provides a convenient tool for studying general boundary conditions of the spin system when coupled to gravity. We are interested in calculating the generating function for the disk amplitude of this model corresponding to the fixed-spin fully magnetised boundary condition, where all of the spins that lie on the boundary have the same orientation.

A given boundary condition is specified by the length of the boundary and a string of labels denoting which spins are present and in which order they are arranged on the boundary. For each Potts spin, $X_i$, this label is a conjugate fugacity, which we denote $x_i$. Choosing an orientation and a marked edge on a given boundary configuration gives a simple way of ordering the boundary labels; starting with the marked edge, go around the boundary via the prescribed orientation. With this association, a natural next step is to treat the fugacities labelling a given boundary configuration as non-commutative variables, reflecting the non-commutativity of the matrices they are conjugate to. Thus, boundary configurations are associated with words generated by the free algebra of these non-commutative variables. The generating function can be expressed as:
\begin{align}
\label{eq2.9}
    \Phi = &\frac{1}{N} \sum_{n=0}^\infty \langle\text{Tr}\, (x_0 X_0 + x_1 X_1 + x_2 X_2)^n\rangle \\
    = & \sum_{w(x_0,x_1,x_2)} w(x_0,x_1,x_2)\langle p_{w(X_0,X_1,X_2)}\rangle,
\end{align}
where
\begin{equation}
\label{eq2.10}
    p_{w(X_0,X_1,X_2)} = \frac{1}{N}\langle \text{Tr}\, w(X_0,X_1,X_2)\rangle,
\end{equation}
represents the disk amplitude for triangulations with a boundary condition specified by the word $w(x_0,x_1,x_2)$. 

Finally, we define special derivative operators, $\Delta_i$, analogous to those defined in \eqref{eq2.7}. Acting from the left, these operate on strings of free variables by annihilating words which start with a different variable from $x_i$, and otherwise removing the leading order $x_i$,
\begin{equation}
\label{eq2.11}
    \Delta_i(x_j f(x_i)) = \delta_{i,j} f(x_i).
\end{equation}
We also define right derivative operators,
\begin{equation}
    \label{eq2.12}
    f(x_i)x_j \overleftarrow{\Delta_i} = \delta_{i,j}f(x_i)
\end{equation}
These derivative operators have the physical interpretation of adding a boundary segment to the generating function corresponding to the associated fugacity $x_i$ \cite{Carroll1996a}.

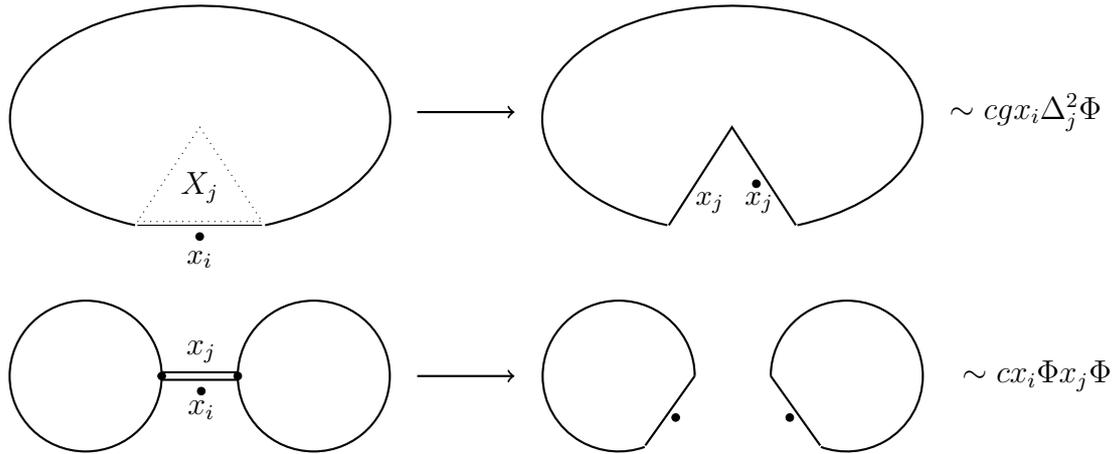
\begin{figure}
\centering
%
\begin{tikzpicture}
\draw[thick] (-3,0) arc (-70:250:2.5cm and 1.5cm);
\draw[dotted] (-4.68,0.05)--(-3.04,0.05) -- (-3.86,1.3) -- (-4.68,0.05);
\node (none) at (-3.86,-0.45) {\small $x_i$};
\node (none) at (-3.86,0.5) {$X_j$};
\draw (-4.68,0) -- (-3.04,0);
\draw [thick,->] (-1,1.5) -- (0.28,1.5);
\draw[thick] (4,0) arc (-70:250:2.5cm and 1.5cm);
\draw[thick] (3.98,0) -- (3.14,1.3) -- (2.31,0);
\draw[fill] (-3.86,-0.15) circle[radius=0.05];
\draw[fill] (3.46,0.55) circle[radius=0.05];
\node (none) at (3.5,0.3) {\small $x_j$};
\node (none) at (2.85,0.3) {\small $x_j$};
\node (none) at (7,1.5) {$\sim cgx_i\Delta_j^2\Phi$};
\draw[thick] (-5.36,-2) circle (1cm);
\draw[thick] (-2.36,-2) circle (1cm);
\draw[fill] (-4.36,-2) circle[radius=0.05];
\draw[fill] (-3.36,-2) circle[radius=0.05];
\draw[thick] (-4.36,-2.05) -- (-3.36,-2.05);
\draw[thick] (-4.36,-1.95) -- (-3.36,-1.95);
\draw[fill] (-3.84,-2.2) circle[radius=0.05];
\node (none) at (-3.84,-2.45) {$x_i$};
\node (none) at (-3.84,-1.675) {$x_j$};
\draw [thick,->] (-1,-2) -- (0.28,-2);
\draw[thick] (3.65,-2) arc (180:-110:1cm and 1cm);
\draw[thick] (2.65,-2) arc (0:290:1cm and 1cm);
\draw[thick] (2,-2.92)--(2.65,-2);
\draw[thick] (4.3,-2.92) -- (3.65,-2);
\draw[fill] (2.4,-2.55) circle[radius=0.05];
\draw[fill] (3.9,-2.55) circle[radius=0.05];
\node (none) at (7.15,-2) {$\sim cx_i\Phi x_j \Phi$};
\end{tikzpicture}
\caption{Result of removing an edge connected to a triangle, or another edge where the marked edge and adjoined component correspond to the different spins.}
\label{Decomp}
\end{figure}

The generating function is constrained by geometric recursive relations. As in the case of pure gravity, the effect of removing an edge of a given triangulation in the generating function can give rise to multiple different triangulations. Two of the four possible outcomes of removing a marked edge are detailed in Fig.\ref{Decomp}. A marked edge corresponding to a spin $X_i$ of a given triangulation can be connected to a triangle within the triangulation, or another edge, which can be associated to either  a different spin $X_j$ or the same spin. By assigning an operation to each of these moves, we can write down the recursion relation, the generating equation, for the generating function.

Consider the first decomposition processes shown in Fig.\ref{Decomp}. By removing an edge connected to a triangle, one generates a diagram where the length of the boundary is increased by one and the new edges have a fugacity conjugate to the spin on the removed triangle. In terms of the original diagram, this new diagram comes with a weight $g$, due to the removed triangle, a weight $c$, due to the propagator from $x_i$ to $x_j$, a factor of $x_j$, due to the new marked edge, and two factors of the derivative operator, $\Delta_j$, due to the removal of an edge of and the addition of two new edges. Collected, this process results in a term $cgx_i\Delta_j^2\Phi$ at the level of the generating function (when $i=j$, there is no factor of $c$).

Now consider the second decomposition process. By removing an edge connected to another edge, we generate a diagram consisting of two new triangulations with the same number of boundary segments and two new marked edges corresponding to the removed edges, accounting for the deleted spins. In terms of the original diagram, we get a factor of $c$, due to the propagator from $x_i$ to $x_j$, a factor $x_i \Phi$ to account for the new diagram with the marked edge $x_i$, and a term $x_j\Phi$ accounting for the second diagram with marked edge $x_j$. Collected, these are translated into $cx_i\Phi x_j\Phi$ at the level of the generating function. When $i=j$ the propagator from $x_i$ to itself is $1$, and so the final term is $x_i\Phi x_i\Phi$.

Collecting all possible deconstruction processes that can occur, we can write down the final generating equation,
\begin{align}
\label{eq2.13}
    \Phi=1 &+ x_0 \Phi\,(x_0 + c x_1 + c x_2)\, \Phi +  g(x_0 + cx_1 +cx_2)\Delta_0^2 \Phi \nonumber \\
    &+ x_1 \Phi\,(cx_0 +  x_1 + cx_2)\, \Phi + g(cx_0+x_1+cx_2)\Delta_1^2 \Phi \\
  &+ x_2 \Phi\,(cx_0 + c x_1 + x_2)\, \Phi + g(cx_0+cx_1+x_2)\Delta_2^2 \Phi \nonumber.
\end{align}
Unfortunately this equation cannot be solved in general to obtain the generating function $\Phi$ for all possible boundary conditions, labelled by words in the free algebra generated by $\{x_0,x_1,x_2\}$. However, we will demonstrate that it can be used to solve for the $\it{fixed}\,\it{spin}$ generating function, which generates disk amplitudes corresponding to triangulations that have a single Potts spin on the boundary.

\section{Computation of Loop Equations}

In this section we derive a closed set of loop equations from the generating equation \eqref{eq2.13}. We first note that by taking derivatives of the generating equation, in the $x_0, x_1$ and $x_2$ directions, we can recast it as follows:
\begin{equation}
\label{eq3.1}
    (1+c)\Delta_0 \Phi + (2c^2- c - 1)(\Phi x_0 \Phi + g \Delta_0^2 \Phi ) = c(\Delta_1 \Phi + \Delta_2 \Phi ).
\end{equation}
This expression is equivalent to the generating equation, but does not determine the leading constant term in $\Phi$. We take this approach because it simplifies the subsequent analysis and makes comparison with Schwinger-Dyson equations clear.  

We can expand the generating function in $x_1,x_2$, giving a series in the fixed spin generating functions and mixed trace amplitudes as functions of $x_0$.
\begin{align}
\label{eq3.2}
    \Phi = & \phi + (x_1 + x_2)\phi_1 + (x_1^2 + x_2^2)\phi_{11} + (x_0 x_1 + x_0 x_2)\Delta_0 \phi_1 + \nonumber \\
    & (x_1 x_2 + x_2 x_1 )\phi_{12} + (x_1^3 + x_2^3)\phi_{111} + (x_1 x_0 x_1 + x_2 x_0 x_2 )\phi_{101} +\\
    & (x_0 x_1 x_1 + x_0 x_2 x_2)\Delta_0 \phi_{11} + (x_0 x_1 x_2 + x_0 x_2 x_1 )\Delta_0 \phi_{12} + \nonumber\\
    & (x_1 x_1 x_2 + x_2 x_2 x_1 )\phi_{112} + (x_1 x_2 x_1 + x_2 x_1 x_2 )\phi_{121} + (x_2 x_1 x_1 + x_1 x_2 x_2 )\phi_{122}+ \cdots \nonumber
\end{align}
where
\begin{equation}
\label{eq3.3}
    \phi_{w(1,2)} = \frac{1}{N}\langle \text{Tr}\,w(X_1,X_2)\frac{1}{1-x_0X_0}\rangle,
\end{equation}
and we use the permutation symmetry of the action to arrange the labels so that the word $w(1,2)$ starts with 1.
In this convention, to isolate a given term of the expansion requires operating on $\Phi$ with the reverse string of derivative operators. For example
\begin{equation}
\label{eq3.4}
    \phi_{112} = \Delta_2\Delta_1\Delta_1\Phi|_{x_1=x_2=0}.
\end{equation}
Furthermore, when applying the derivative operator on the right hand side to functions with cyclic symmetry (such as $\Phi$) it is equivalent to concatenating the right-hand set of operators with the operators on the left-hand side, but now as right-acting operators. For example, 
\begin{equation}
\label{eq3.5}
    \Delta_1\Delta_0\Phi\overleftarrow{\Delta_1\Delta_2} = \Delta_1\Delta_0\Delta_1\Delta_2\Phi.
\end{equation}
This equivalence can be seen by applying this rule to the series expansion of the generating function. It fails, however, for terms quadratic in the generating function, since they are not cyclic.

By collecting each term in the expansion of the generating function within the generating equation, we can identify constraints for the fixed spin amplitude. There is no known way of demonstrating that any set of constraints we construct results in a closed system of equations, and in principle we can continue to generate constraints that we can't solve to find the fixed spin amplitude. 

We will proceed by computing a set of loop equations in the variable $x_0$. We label these by a corresponding string of variables in the expansion of the generating equation. For example the loop equation labelled by $w(x_1,x_2)x_0\cdots x_0 v(x_1,x_2)$, where $w(x_1,x_2)$ and $v(x_1,x_2)$ are arbitrary words in $x_1$ and $x_2$, will correspond to operating on the generating function with $w(\Delta_1,\Delta_2)$ on the left and $v(\Delta_1,\Delta_2)$ on the right and setting $x_1$ and $x_2$ to zero, thereby producing an equation in $x_0$ only.
\subsection{$x_0\cdots x_0$}
The leading order term in \eqref{eq3.1}, after taking $x_1,x_2\rightarrow 0$ is
\begin{equation}
\label{eq3.6}
    (1+c)\Delta_0\phi -(1+c-2c^2)(x_0 \phi^2 + g \Delta_0^2\phi) = 2c\phi_1.
\end{equation}
\subsection{$x_1x_0\cdots x_0x_1$}
Following this we can extract the loop equation corresponding to the string $x_1x_0\cdots x_0 x_1$ by operating on the left and right hand side of the generating equation with $\Delta_1$ and setting $x_1,x_2\rightarrow0$:
\begin{equation}
\label{eq3.7}
    (1+c)\phi_{101}-(1+c-2c^2)(x_0\phi_1^2+g\phi_{1001})=c(\phi_{111}+\phi_{121}).
\end{equation}
\subsection{$x_1x_0\cdots $}
We can also operate with $\Delta_1$ on the left hand side only. However, this operation is not Hermitian, in the sense that the resulting equation extracted does not necessarily contain expectation values of Hermitian matrices. This is fixed by adding the conjugate operation, which is where we operate separately with $\overleftarrow{\Delta_1}$ on the right hand side, and add the contributions together:
\begin{equation}
\label{eq3.8}
    (1+c)\Delta_0\phi_1 -(1+c-2c^2)(x_0\phi\phi_1 + g\Delta_0^2\phi_1) = c (\phi_{11}+\phi_{12}).
\end{equation}
\subsection{$x_2x_1x_0\cdots $}
We can then look at mixed strings of the form $x_2x_1x_0\cdots$ (and the conjugate string):
\begin{equation}
\label{eq3.9}
(1+c)\Delta_0\phi_{12} - (1+c-2c^2)(x_0\phi\phi_{12}+g\Delta_0^2\phi_{12}) = c(\phi_{121}+ \phi_{(122)}),
\end{equation}
where the brackets in the subscript mean symmetrisation between the string and the reversed string:
\begin{equation}
\label{eq3.10}
    \phi_{(122)} = \frac{1}{2}(\phi_{122}+\phi_{221}).
\end{equation}
Crucially, the fact that terms like $\phi_{122}$ and $\phi_{221}$ do not arise independently, but in the symmetrised combination above means we have a smaller set of independent functions.

\subsection{$x_1x_2x_1x_0\cdots$}
\begin{equation}
\label{eq3.11}
    (1+c)\Delta_0\phi_{121} - (1+c-2c^2)(x_0\phi\phi_{121}+g\Delta_0^2\phi_{121}) = c(\phi_{1212}+\phi_{(1121)}).
\end{equation}
\subsection{$x_1x_0x_2x_0\cdots$}
\begin{equation}
\label{eq3.12}
    (1+c)\Delta_0\phi_{102} - (1+c-2c^2)(x_0\phi\phi_{102}+p_1 \phi_1+g\Delta_0^2\phi_{102})=c(\phi_{(1102)}+\phi_{(1201)}).
\end{equation}
\subsection{$x_1x_0x_1x_0\cdots$}
\begin{equation}
\label{eq3.13}
    (1+c)\Delta_0\phi_{101} -(1+c-2c^2)(x_0\phi\phi_{101}+p_1\phi_1 + g \Delta_0^2\phi_{101}) = c(\phi_{(1101)}+\phi_{(1202)}).
\end{equation}
\subsection{$x_0x_1x_2x_0\cdots$}
\begin{equation}
\label{eq3.14}
    (1+c)\Delta_0^2\phi_{12}-(1+c-2c^2)(x_0\phi\Delta_0\phi_{12}+\phi_{12} + g \Delta_0^3\phi_{12}) = c(\phi_{(1201)}+\phi_{(1202)}).
\end{equation}
\subsection{$x_1x_2x_0\cdots x_0 $}
\begin{equation}
\label{eq3.15}
    (1+c)\Delta_0^2\phi_{12}-(1+c-2c^2)(x_0(\Delta_0\phi)\phi_{12}+\phi_{12}+g\Delta_0^3\phi_{12}) = c(\Delta_0\phi_{(112)}+\Delta_0\phi_{121}).
\end{equation}

At each level of expansion it appears that we have new objects appearing, implying that we can never arrive at a closed system of equations for the resolvent. However, we can appeal to the permutation symmetry of the action once more, and construct loop equations by expanding around another variable, say, $x_2$. By permutation symmetry, any loop equation for the correlators of resolvents in $x_2$ and a word in $X_0,X_1$, can always be expressed as a loop equation for the original resolvent in $x_0$.
\subsection{$x_2\cdots x_2$}
First, by expanding the generating equation in $x_2$ and setting $x_0,x_1\rightarrow0$, and using permutation symmetry, we find the following loop equation:
\begin{equation}
\label{eq3.16}
    \phi_1 - (1+c-2c^2)g\phi_{11} = c\Delta_0\phi.
\end{equation}
We now list a set of constraints using this procedure.
\subsection{$x_1x_2\cdots x_2 x_1$}
\begin{equation}
\label{eq3.17}
    (1+c)\phi_{121} - (1+c-2c^2)g\phi_{1221} = c(\phi_{111} + \phi_{101}).
\end{equation}
\subsection{$x_1x_2\cdots$}
\begin{equation}
\label{eq3.18}
    (1+c)\phi_{12}-(1+c-2c^2)g\phi_{(112)} = c(\phi_{11}+\Delta_0\phi_{1}).
\end{equation}
\subsection{$x_0x_2\cdots$}
\begin{equation}
\label{eq3.19}
    (1+c)\phi_{11} - (1+c-2c^2)(\phi+g\phi_{111}) = c(\phi_{12}+\Delta_0\phi_1).
\end{equation}
\subsection{$x_1x_1x_2\cdots$}
\begin{equation}
\label{eq3.20}
    (1+c)\phi_{(112)}-(1+c-2c^2)g\phi_{1122} = c(\phi_{111}+\Delta_0\phi_{11}).
\end{equation}
\subsection{$x_2x_1x_2\cdots$}
\begin{equation}
\label{eq3.21}
    (1+c)\phi_{102} - (1+c-2c^2)g\phi_{(1102)} = c(\phi_{101} + \Delta_0^2\phi_1).
\end{equation}
\subsection{$x_2x_0x_2\cdots$}
\begin{equation}
\label{eq3.22}
    (1+c)\phi_{101} - (1+c-2c^2)(p_1\phi + g \phi_{(1101)}) = c(\phi_{102} + \Delta_0^2\phi_1).
\end{equation}
\subsection{$x_1x_0x_2\cdots$}
\begin{equation}
\label{eq3.23}
    (1+c)\phi_{121} - (1+c-2c^2)(p_1\phi + g\phi_{(1121)}) = c(\phi_{(112)}+\Delta_0\phi_{12}).
\end{equation}
\subsection{$x_0x_1x_2\cdots x_2 x_0$}
\begin{equation}
\label{eq3.24}
    (1+c)\phi_{(1222)} - (1+c-2c^2)(2\phi_{12}+g\phi_{(12222)}) = c(\phi_{1212} + \phi_{(1202)}).
\end{equation}
\subsection{$x_1x_2\cdots x_2 x_0x_0$}
\begin{equation}
\label{eq3.25}
    (1+c)\phi_{(1222)}-(1+c-2c^2)(\phi_{12}+p_1\phi_1 + g\phi_{(12222)}) = c(\phi_{1122}+\phi_{(1102)}).
\end{equation}
\subsection{$x_0x_2\cdots x_2x_0x_2$}
\begin{equation}
\label{eq3.26}
    (1+c)\phi_{(1211)}-(1+c-2c^2)(\phi_{12}+p_1\phi_1+g\phi_{(12111)})=c(\phi_{1001}+\phi_{(1201)}).
\end{equation}
\subsection{$x_0x_2x_0x_2\cdots$}
\begin{equation}
\label{eq3.27}
    (1+c)\phi_{(1211)}-(1+c-2c^2)(\phi_{12}+p_{12}\phi+g\phi_{(12111)})=c(\phi_{(1202)}+\Delta_0\phi_{101}).
\end{equation}
\subsection{$x_0x_2\cdots x_2 x_1 x_0$}
\begin{equation}
\label{eq3.28}
    (1+c)\phi_{(1222)}-(1+c-2c^2)(2\phi_{12}+g\phi_{(12222)})=c(\phi_{1212}+\phi_{(1202)}).
\end{equation}
\subsection{$x_0x_0x_1x_2\cdots$}
\begin{equation}
\label{eq3.29}
    (1+c)\phi_{(1222)}-(1+c-2c^2)(\phi_{12}+p_1\phi_1 + g\phi_{(12222)})=c(\phi_{1221}+\Delta_0\phi_{(112)}).
\end{equation}
This forms a closed system of equations and we can solve them to find an algebraic curve describing $\phi$.

\section{The Spectral Curve} 

In this section, we outline how the loop equations we have computed can be used to write down a single algebraic equation defining the one-loop function, which we call the `spectral curve'. We first note that $\phi_1$ is written as a function of $x_0$ and $\phi$ in \eqref{eq3.6}. Using \eqref{eq3.16} and \eqref{eq3.6} we can also write $\phi_{11}$ in terms of $x_0,\phi$. We can continue this successively for $\phi_{12}$ using \eqref{eq3.7} (and the previous expressions for $\phi_1, \phi_{11}$ in terms of $x_0,\phi$), $\phi_{(112)}$ using \eqref{eq3.18}, $\phi_{121}$ using \eqref{eq3.9}, $\phi_{111}$ using \eqref{eq3.19}, $\phi_{1122}$ using \eqref{eq3.20} and $\phi_{(1121)}$ using \eqref{eq3.23}. 

We can then construct an algebraic expression in terms of the above variables using the remaining equations. We will briefly outline one way of doing this. By taking the remaining equations that have not already been used we can construct several independent equations by eliminating suitable functions. Eventually a single equation can be derived relating the functions previously described, $\phi_1, \phi_{11}, \phi_{12}, \phi_{(112)},\phi_{121}, \phi_{111}, \phi_{1122}$ and $\phi_{(1121)}$. Using the expressions for these in terms of $\phi$ and $x_0$ leaves us with a single algebraic equation for $\phi$. 

This final algebraic equation contains a number of undetermined constants that arise from the derivative operation $\Delta_0$ applied to various functions. These are $p_1, p_{12}, p_{11}, p_{112}, p_{012}, p_{1212}$ and $p_{0121}$. Fortunately these coefficients are not all independent, and there are recurrence relations between these correlators as well, which result in a rather non-trivial cancellation of terms that simplify the final algebraic equation and reduces its order to 5. For example, the leading order term in the series expansion of \eqref{eq3.6} in $x_0$, gives
\begin{equation}
    \label{eq4.1}
    g(1+c-2c^2)p_{11} = (1-c)p_1.
\end{equation}
Similarly, the first order term in the expansion of \eqref{eq3.16} gives
\begin{equation}
    \label{eq4.2}
    p_{12} - g(1+c-2c^2)p_{112} = cp_{11},
\end{equation}
and the zeroth order term of \eqref{eq3.24}-\eqref{eq3.25} gives the following constraint,
\begin{equation}
    \label{eq4.3}
    c(p_{1212} + p_{0121} - p_{1122} - p_{1120}) = -(1+c-2c^2)(p_{12} - p_1^2).
\end{equation}
$p_{1122}, p_{1120}$ can be expressed in terms of $p_{012}$, $p_{12}$ and $p_1$, so the total number of undetermined constants is actually 4.

Finally, we can write down the spectral curve, the algebraic curve describing the generating function, as the following quintic
\begin{equation}
    \label{eq4.4}
    f_5 y^5 + f_4 y^4 + f_3 y^3 + f_2 y^2 + f_1 y + f_0 = 0
\end{equation}
where we use the shifted function, $y$, given by
\begin{equation}
    \label{eq4.5}
    \phi = -\frac{y}{x} - \frac{g}{x^3} + \frac{1}{(1-c)} \frac{1}{x}
\end{equation}
and the coefficients are functions of $x$, given by the following expressions
\begingroup
\allowdisplaybreaks
\begin{align}
    \label{eq4.6}
   f_5 = & -4 (c-1)^8 g^3 (2 c x+x)^6 \\
   \label{eq4.7}
   f_4 = & -(-1 + c)^6 g^2 (x + 2 c x)^4 (4 (1 + c - 2 c^2)^2 g^2 + 4 (-1 + c) (1 + 2 c) (1 + 5 c) g x \nonumber
   \\ &  - c (18 + 19 c) x^2)  \\
   \label{eq4.8}
   f_3 = & 2 (c-1)^4 (2 c+1)^2 g x^3 \bigg(2 (-2 c^2+c+1)^4 g^3 p_1 x^3 + 6 c (-2 c^2+c+1)^3 g^3 \nonumber
   \\ & +x^3 \left((c-1)^3 (2 c+1)^3 (3 c+2) g^2-c^2 (c (12 c+31)+13)\right)
   \\ & +(c-1) (2 c+1) g x^2 \left(3 c (c (17 c+16)+3)-2 \left(-2 c^2+c+1\right)^3 g^2\right) \nonumber
   \\ & -3 c (3 c-1) \left(-2 c^2+c+1\right)^2 g^2 x \bigg) \nonumber \\
    \label{eq4.9}
   f_2 = & -c \left(-2 c^2+c+1\right)^2 x^2 \bigg(2 (c-1)^4 (2 c+1)^2 g^3 x^4 \big(4 (c-1) (2 c+1) g p_{\{1,2\}} \nonumber
    \\ & +(12 c+13) p_1\big)+2 x^4 \left((c-1)^3 (2 c+1) (4 c (2 c+5)+9) g^2-6 c^2\right) \nonumber
    \\ & -(c-1)^2 g^2 x^2 \left(3 c (c (53 c+38)+5)-10 \left(-2 c^2+c+1\right)^3 g^2\right)
    \\ & +13 (c-1)^4 c (2 c+1)^2 g^4-4 (c-1)^3 c (2 c+1) (8 c+7) g^3 x \nonumber
    \\ & +2 (c-1) g x^3 \left(2 (c+2) (c-1)^3 (2 c g+g)^2+c (c (43 c+52)+13)\right)\bigg) \nonumber \\
    \label{eq4.10}
   f_1 = & -2 (c-1) c (2 c+1)^2 x \bigg((c-1)^3 (2 c+1) g^2 x^5 \big(2 (c-1) c g  \nonumber
   \\ & \left(2 \left(-2 c^2+c+1\right) g p_{\{0,1,2\}}-3 (4 c+3) p_{\{1,2\}}\right) \nonumber
   \\ & +p_1 \left((c-1)^3 (2 c g+g)^2-9 c (c+3)\right)\big)+3 (c-1)^4 c^2 (2 c+1) g^4 \nonumber
   \\ & -3 (c-1)^3 c^2 (7 c+4) g^3 x \nonumber
   \\ & +x^4 \left(6 c^2 (2 c+1) (c-1)^3 g^2-6 c^2 (c+1)+(2 c+1)^3 (c-1)^6 g^4\right)
   \\ & +3 (c-1) c g x^3 \left((2 c+1) (5 c+3) (c-1)^3 g^2+c (13 c+5)\right) \nonumber
   \\ & -2 (c-1)^2 c g^2 x^2 \left(2 (c-1)^3 (2 c g+g)^2-3 c\right) \nonumber
   \\ & +(c-1)^2 g x^5 \left((c+1) (c-1)^3 (2 c g+g)^2+c (c (7 c-18)-13)\right)\bigg) \nonumber \\
   \label{eq4.11}
   f_0 = & -c^2 \bigg((c-1)^2 g x^6 \big(4 (c-1) (2 c+1) g  \nonumber
   \\ & \Big(\Big(\left(-2 c^2+c+1\right)^3 g^2+c (c (18 c+31)+9)\Big) p_{\{1,2\}} \nonumber
   \\ & +(c-1) c (2 c+1) g \left(2 (c-1) (2 c+1) g p_{\{1,2,0,2\}}+(8 c+5) p_{\{0,1,2\}}\right)\Big) \nonumber
   \\ & +3 \left(-2 c^2+c+1\right)^4 g^3 p_1^2 \nonumber
   \\ & +2 p_1 \left((3 c+2) \left(-2 c^2+c+1\right)^3 g^2+2 c (9-4 c ((c-10) c-9))\right)\big) \nonumber
   \\ & +(2 c+1)^2 \Big((c-1)^4 c^2 g^4-6 (c-1)^3 c^2 g^3 x
   \\ & +x^4 \left(-12 c^2+(2 c+1)^2 (c-1)^6 g^4-6 c (2 c+1) (c-1)^3 g^2\right) \nonumber
   \\ & -(c-1)^2 x^6 \left((c-1)^2 (c (c+8)+3) g^2+12 c\right) \nonumber
   \\ & -2 (c-1)^2 (2 c+1) g x^5 \left((2 c+1) (c-1)^3 g^2+2 c\right) \nonumber
   \\ & +4 (c-1) c g x^3 \left(2 (2 c+1) (c-1)^3 g^2+c\right) \nonumber
   \\ & -(c-1)^2 c g^2 x^2 \left(2 (c-1)^3 (2 c+1) g^2-9 c\right)\Big)\bigg) \nonumber
\end{align}
\endgroup
This can be shown to agree with past computations of the spectral curve \cite{Eynard1999,Atkin2016,Zinn-Justin2000}.

\section{Conclusion}

In this note we have addressed the long standing problem of computing the fixed spin generating function for the disk amplitude of the 3-state Potts model coupled to 2D quantum gravity, completing the calculation of \cite{Carroll1996a}. We derived the generating equation for the full generating function, developed the method of constructing the loop equation constraints in this geometric setup, and computed a finite set of independent loop equations that could then be solved to write down the spectral curve, in agreement with previous derivations \cite{Eynard1999,Atkin2016,Zinn-Justin2000}. This method presents a number of advantages over the usual approach of asserting reparameterisation invariance of the matrix integral definition of the model, and the saddle point method, being conceptually cleaner with a clear relationship between the combinatorics of random planar maps and associated loop equations. It also provides a systematic method of constructing disk amplitudes with boundary insertions once the fixed spin generating function is computed, such as $\phi_{1}(x)$ which generates planar maps with a single $x_1$ spin insertion with a boundary otherwise composed of $x_0$ spins.

This note also presents a number of opportunities for further research. For example, it would be interesting to see whether we could exploit the method further and push beyond the planar limit to calculate the generating function at higher genus. It would also be interesting to see whether a topological recursion can be developed \cite{Eynard2007}.

Exploring whether this method could be used to calculate more general boundary conditions would be an interesting avenue of research. Calculating the mixed and free boundary conditions in the 3-state Potts model was a question recently addressed in \cite{Atkin2015,Atkin2016} using the saddle-point method. Doing this same calculation in the loop equation framework used here would provide a concrete combinatoric argument to the spectral curves derived therein, but this remains an open question. Furthermore, this method was originally used to calculate the non-integrable interpolating boundary generating function for the Ising model coupled to 2D gravity \cite{Carroll1998} and it may be possible to calculate a similar object for the 3-state Potts model. Lastly, they were able to calculate the boundary generating functions for the Kramers-Wannier dual to the Ising model \cite{Carroll1996}. A new boundary condition was recently derived in \cite{Kulanthaivelu2019} for the 3-state Potts model, and it was conjectured that this corresponds to the gravitational analogue of the `new' boundary condition for the 3-state Potts model \cite{Affleck1998}. To prove this conjecture one could calculate the dual theory and compute the mixed spin amplitudes, which should be equivalent to the `new' boundary condition in the scaling limit \cite{Fuchs1998}. However this model is a mixed complex-Hermitian matrix model whose solution is currently out of reach. It would be exciting to see whether the method used in this paper could be adapted to address this question. 

\ack
This work is supported by the  STFC grant ST/N504233/1. I thank John Wheater for useful discussions and
constant encouragement.

\appendix

\section{Loop Equations via Direct Reparameterisations}
One can derive loop equations for \eqref{eq1.1} by demanding reparameterisation invariance for the underlying matrix degrees of freedom of the model. This method is slightly less general than the combinatorial method developed in this paper, but is functionally identical to the approach taken here, in that the constraints derived here are a subset of those that can be constructed using the generating equation.

Under an infinitesimal reparameterisation of one of the matrix degrees of freedom $X_i\rightarrow \bar{X_i}=X_i+\epsilon \delta X_i$, we have
\begin{equation}
    Z\rightarrow \bar{Z} = \int \prod_i [d X_i][1+\epsilon J]e^{-N S(\{X_i\})}[1-\epsilon K],
\end{equation}
where the $J$ is the first order term in the $\epsilon$-expansion of the Jacobian, coming from the the variation of the measure, and $K$ comes from the variation of the action. The variation of the action has a simple expression,
\begin{equation}
    K = N \text{Tr}\delta X_i S'(\{X_i\}).
\end{equation}
Whereas the infinitesimal variation of the Jacobian can be calculated using the so-called `split' and `merge' rules:
\begin{itemize}
    \item a reparameterisation of the form $\delta X_i = A \frac{1}{z-X_i} B$ generates a Jacobian,
    \begin{equation}
        J = \text{Tr}\bigg(A\frac{1}{z-X_i}\bigg)\text{Tr}\bigg(\frac{1}{z-X_i}B\bigg) + \text{contributions from A and B},
    \end{equation}
    \item a reparameterisation of the form $\delta X_i = A \text{Tr}\bigg(B\frac{1}{z-X_i}\bigg)$ generate a Jacobian,
    \begin{equation}
        J = \text{Tr}\bigg(A\frac{1}{z-X_i}B\frac{1}{z-X_i}\bigg) + \text{contributions from A and B},
    \end{equation}
    \item variations due to A and B are computed recursively via the chain rule.
\end{itemize}

This method was first proposed in \cite{Bonnet1999}, but we go beyond and give explicitly a complete set of reparameterisations that result in a closed set of constraints on the fixed spin generating function. These reparameterisations, following the order given in \eqref{eq3.6}-\eqref{eq3.29}, are

\begin{enumerate}
    \item $\delta X_0 = \frac{1}{z- X_0}$
    \item $\delta X_0=  X_1\frac{1}{z- X_0} X_1$
    \item $\delta X_0 =  X_1 \frac{1}{z- X_0}+\frac{1}{z- X_0} X_1$
    \item $\delta  X_0 =  X_2  X_1 \frac{1}{z- X_0} + \frac{1}{z- X_0} X_1 X_2$
    \item $\delta  X_0 =  X_1  X_2  X_1 \frac{1}{z- X_0}+\frac{1}{z- X_0} X_1  X_2  X_1$
    \item $\delta X_0 =  X_1 X_0 X_2\frac{1}{z- X_0}+\frac{1}{z- X_0} X_2 X_0 X_1$
    \item $\delta X_0 =  X_1 X_0 X_1\frac{1}{z- X_0}+\frac{1}{z- X_0} X_1 X_0 X_1$
    \item $\delta X_0 =  X_0 X_1 X_2\frac{1}{z- X_0} +\frac{1}{z- X_0} X_2 X_1 X_0$
    \item $\delta X_0 =   X_1 X_2\frac{1}{z- X_0} X_0+ X_0\frac{1}{z- X_0} X_2 X_1$
    \item $\delta  X_0 = \frac{1}{z- X_2}$
    \item $\delta X_0 =  X_1\frac{1}{z- X_2} X_1$
    \item $\delta  X_0 =  X_1 \frac{1}{z- X_2}+\frac{1}{z- X_2} X_1$
    \item $\delta X_0 = \frac{1}{z- X_2} X_0 +  X_0\frac{1}{z- X_2}$
    \item $\delta X_0 = \frac{1}{z- X_2} X_1^2+ X_1^2\frac{1}{z- X_2}$
    \item $\delta X_0 =  X_2 X_1\frac{1}{z- X_2}+\frac{1}{z- X_2} X_1 X_2$
    \item $\delta X_0 =  X_2 X_0\frac{1}{z- X_2}+\frac{1}{z- X_2} X_0 X_2$
    \item $\delta  X_0 =  X_1  X_0 \frac{1}{z- X_2}+\frac{1}{z- X_2} X_0 X_1$
    \item $\delta X_0 =  X_0  X_1 \frac{1}{z- X_2} X_0  + X_0 \frac{1}{z- X_2} X_1 X_0$
    \item $\delta X_0 =  X_1 \frac{1}{z- X_2} X_0^2 +  X_0^2 \frac{1}{z- X_2} X_1$
    \item $\delta  X_0 =  X_0 \frac{1}{z- X_2} X_0  X_2 +  X_2  X_0 \frac{1}{z- X_2} X_0$
    \item $\delta  X_0 =  \frac{1}{z- X_2} X_0  X_2  X_0 +  X_0  X_2  X_0 \frac{1}{z- X_2}$
    \item $\delta X_0 =  X_0\frac{1}{z- X_2} X_1 X_0+  X_0 X_1\frac{1}{z- X_2} X_0$
    \item $\delta X_0 =  X_0^2 X_1\frac{1}{z- X_2}+\frac{1}{z- X_2} X_1 X_0^2$
\end{enumerate}


\section*{References}

\bibliography{ref2.bib}

\end{document}